\documentclass[twocolumn]{aastex7}

\usepackage{subcaption}
\usepackage{threeparttable}
\usepackage{gensymb}

\usepackage[T1]{fontenc}
\usepackage{romanbar}
\usepackage{hyperref} 
\usepackage{graphicx}
\usepackage{color}
\usepackage{amsmath}
\usepackage{pgf-pie}
\usepackage{float}

\shortauthors{Huseby et al.}

\begin{document}

\title{Ultraviolet Radiation Effects on the Optical Properties of Water-Dominated Exoplanet Hazes}

\author[0009-0009-5477-9375]{Lori Huseby}
\affiliation{Lunar and Planetary Laboratory, University of Arizona, Tucson, AZ 85721, USA}
\email{lhuseby@arizona.edu} 

\author[0000-0002-6721-3284]{Sarah E. Moran}
\altaffiliation{NHFP Sagan Fellow}
\affiliation{Space Telescope Science Institute, Baltimore, MD 21218, USA}

\email{samoran@stsci.edu}

\author[0000-0003-3759-9080]{Tiffany Kataria}
\affiliation{Jet Propulsion Laboratory, Pasadena, CA 91011, USA}
\email{Tiffany.Kataria@jpl.nasa.gov}

\author[0000-0002-5251-2943]{Mark S. Marley}
\affiliation{Lunar and Planetary Laboratory, University of Arizona, Tucson, AZ 85721, USA}
\email{marksmarley@arizona.edu}

\author[0000-0002-6694-0965]{Chao He}
\affiliation{School of Earth and Space Sciences, University of Science and Technology of China, Hefei, People's Republic of China}
\affiliation{Department of Earth and Planetary Sciences, Johns Hopkins University, Baltimore, MD 21218, USA}
\email{chaohe23@ustc.edu.cn}

\author[0009-0007-6910-6347]{Cara Pesciotta}
\affiliation{Department of Earth and Planetary Sciences, Johns Hopkins University, Baltimore, MD 21218, USA}
\email{cpescio1@jhu.edu}

\author[0000-0003-4596-0702]{Sarah M. Hörst}
\affiliation{Department of Earth and Planetary Sciences, Johns Hopkins University, Baltimore, MD 21218, USA}
\email{sarah.horst@jhu.edu}

\author[0000-0002-0183-1581]{Neil Pearson}
\affiliation{Lunar and Planetary Laboratory, University of Arizona, Tucson, AZ 85721, USA}
\affiliation{Planetary Science Institute, Tucson, AZ 85719, USA}
\email{npearson@psi.edu}



\author[0000-0002-7743-3491]{Vishnu Reddy}
\affiliation{Lunar and Planetary Laboratory, University of Arizona, Tucson, AZ 85721, USA}
\email{vishnureddy@arizona.edu}

\author[0000-0002-8507-1304]{Nikole K. Lewis}
\affiliation{Department of Astronomy, Cornell University, Ithaca, NY 14853, USA}
\email{nkl35@cornell.edu}

\author[0000-0001-7273-1898]{V{\'e}ronique Vuitton}
\affiliation{Univ. Grenoble Alpes, CNRS, IPAG, 38000 Grenoble, France}
\email{Veronique.Vuitton@univ-grenoble-alpes.fr}

\correspondingauthor{lhuseby@arizona.edu}
\correspondingauthor{chaohe23@ustc.edu.cn}

\begin{abstract}
Temperate sub-Neptune and terrestrial exoplanets could contain large inventories of water in various phases, such as water-dominated atmospheres or even oceans. 
Observations have shown that many exoplanets, including water worlds, likely contain photochemically-generated hazes. 
Haze particles are a key source of organic matter and may impact the evolution or origin of life; their optical properties are imperative for interpreting observations through theoretical atmospheric modeling.
Modelers have thus far assumed haze optical properties that may not represent hazes under sub-Neptune and terrestrial atmospheric conditions. 
Often orbiting close to M-dwarf stars, these planets receive large amounts of radiation, especially during flaring events, which may accelerate atmospheric escape and affect atmospheric compositions. 
Here, we present optical constants of experimentally-generated sub-Neptune haze analogs before and after UV irradiation across a broad wavelength range (0.5 to 8 $\mu$m). 
We find that UV-irradiation alters haze optical constants which become generally more absorbing in this wavelength range, which we hypothesize is due to our sample containing more oxygen-rich absorbing bands post irradiation.
We use Virga and PICASO to simulate transmission spectra of potentially hazy water-dominated planets GJ 1214b and LHS 1140b, accounting for irradiated haze layers in their atmospheres. 
For our GJ1214b CH$_4$-rich haze modeled case, we see a difference in the N-H feature at 2.6$\mu$m in the resulting transmission spectrum between irradiated and unaltered haze that should be observable within current JWST capabilities. 
Broadly, we demonstrate the importance of using more representative optical constants, as they have an impact on current and future atmospheric composition interpretations.

\end{abstract}

\section{\textbf{Introduction}}

Due to their high planet-star contrast ratio and transit probability, cooler sub-Neptune (T$_{\rm eff}$ < 1000\,K, 2.0 -- 4.0 R$_{\rm{Earth}}$) \cite[]{Borucki2011Populations,Fressin2013OccurenceRates,Fulton2017Gap} exoplanets that orbit M-dwarf stars are optimal targets for observational atmospheric composition characterization (e.g., \citealt{2019HenryMdwarf}).
However, clouds and hazes make exoplanet transmission spectra interpretation more difficult by dampening spectral features in infrared wavelengths and causing larger than Rayleigh scattering slopes in optical wavelengths (e.g., \citealt{Ohno2020Rayleigh}), which can limit our ability to infer atmospheric compositions of these worlds (e.g., \citealt{Morley2013Clouds1214,Kreidberg20141214b,Knutson2014GJ436b,Knutson2014HD97658b,Dragomir2015GJ3470b,LibbyRoberts2020Superpuff,Kreidberg2022HD106315c}). 
Hazes, in particular, vary widely in composition, shape, and size, which change the resulting optical properties of the particles and lead to variations in their opacities \cite[]{Arney2016PaleOrangeDot,Arney2017PaleOrangeDot,Gavilan2018Haze,Corrales2023HazesCO,He2023Methods}. 

Optical constants, defined as the real ($n$) and imaginary ($k$) refractive indices, quantify how light interacts with a material, and help infer the composition of the worlds being observed. 
Optical constants allow us to transition from laboratory measurements of our hazes into values that can be input into modeling codes to generate exoplanet transmission spectra.
Several studies report the optical properties of hazes formed under different conditions and compositions (e.g., \citealt{Khare1984OpticalProp,Lavvas2017HazeUncertainty,He2022Titan,Corrales2023HazesCO,He2023Methods,Drant2024opticalprop,Li2025Haze,Drant2026setups, Pesciotta2026Haze}). 
Overall, the resulting optical constants of these laboratory studies are diverse. 
Specifically, \cite{He2023Methods} measured optical constants of hazes on 1000x solar metallicity atmospheres, representing water-rich atmospheres under two different temperatures and compositions, and found that the differences in their resulting haze optical constants had a detectable effect on exoplanet atmosphere transmission spectra. 
Exoplanet aerosol models have been used to interpret a variety of observations of exoplanet atmospheres and also predict future observations. 
In particular, many studies have focused on explaining observations of individual planets with aerosol models (e.g., \citealt{Kreidberg20141214b,Knutson2014GJ436b,Benneke2024TOI270d}). 
Recently, \citet{Jaziri2025k218haze} showed that using different haze optical constants in retrievals for planet K2-18b influence the retrieved parameters due to IR haze features.
However, these comparisons often run into degeneracies between potential hazes and different atmospheric compositions, and it is unknown how representative their conclusions are to any given exoplanet due to a lack of accurate optical properties of their hazes. 
As more exoplanet atmospheres are characterized, distinct haze optical constants representative of relevant atmospheric conditions are necessary for modeling their atmospheres correctly. 

With the complications hazes produce for exoplanet observations, it is also important to understand how stellar activity on M-dwarfs affect the evolution of sub-Neptune or terrestrial planet atmospheres.
These close-in planets can be dramatically affected by stellar activity and, more specifically, stellar flares (e.g., \citealt{Mordasini2009WaterWorld,Howard2020FlareHab,doAmaral2022Flare,Konings2022FlaresMdwarfs,doAmaral2025flares}).
It is critical to understand the properties of such haze particles in this high energy environment due to the large amount of irradiation they receive.
Haze compositional changes during a stellar flare will affect the resulting aerosol optical constants, but the extent to which this change can be observed is still unknown.
So far, haze interactions with UV light representing flares have only been studied for Titan and Earth-like haze analogs \citep{Gavilan2018Haze,Carrasco2018VUVhaze}. 
These works found that the total absorptivity of the hazes decreased over the irradiation period, losing hydrogen-carrying molecules and increasing pi bonds to create more complex molecules. 
Neither of these works provide changes in optical constants pre- and post-irradiation, nor are they representative of potential water-dominated exoplanet atmospheres. 
In our previous work \citep{Huseby2025Hazes}, we showed that the spectral features of water-rich atmospheric hazes before and after UV-irradiation change significantly, suggesting that they are compositionally distinct hazes post-irradiation. 
Here, we derive the optical properties, including the real and imaginary refractive indices, of our samples pre- and post-irradiation. We input our newly derived optical properties into simulated transmission spectra of water-dominated atmospheres.
This will demonstrate whether the amplitude of spectral changes will be observable in real exoplanetary transmission spectra.

In Section \ref{sec:Methods}, we briefly describe our experimental methods including the haze sample production (\ref{subsec:hazeproduction}) and haze UV bombardment process (\ref{subsec:FTIR}), as well as the derivation of optical constants that represents the bulk of this new analysis (\ref{subsec:Opticalcalculation}). 
Section \ref{sec:Results} describes the optical constant results of both samples pre- and post-irradiation. 
In Section \ref{sec:TransmissionModel}, we discuss the implications of our results on modeled transmission spectra of GJ 1214b, which shows signs of high-altitude aerosols \citep{Kempton2023GJ1214,Gao2023GJ1214}, and LHS 1140b, which is a potentially habitable water world \citep{Damiano2024LHS1140b}, and their importance to the wider exoplanet science community.

\section{\textbf{Experimental Methods}} \label{sec:Methods}
\subsection{Haze Analog Production}\label{subsec:hazeproduction}
\begin{figure}[h]
    \begin{subfigure}[b]{\columnwidth}
        \begin{tikzpicture}
        \pie[color = {blue!50, pink!50,orange!70,red!100}, radius=1.5]{75/H$_{2}$O, 10/N$_{2}$, 10/CO$_{2}$, 5/\textbf{CH$_{4}$}}
        \pie[color = {blue!50, pink!50,orange!70,green!100}, radius=1.5, pos={4,0}]{75/H$_{2}$O, 10/N$_{2}$, 10/CO$_{2}$, 5/\textbf{CO}}
        \end{tikzpicture}
        \caption{Gas mixtures by mixing ratio used in \citet{Huseby2025Hazes} and this work.}
        \label{fig:GasPieCharts_Huseby2025}
    \end{subfigure}

    \begin{subfigure}[b]{\columnwidth}
        \begin{tikzpicture}
        \pie[color = {pink!50,red!100}, radius=1, hide number,pos={0,0}]{90/, 10/}
        \pie[color = {blue!50, yellow!60, pink!50,orange!70,red!100}, radius=1, hide number, pos={2.2,0}]{66/, 16/, 6.5/, 4.9/, 6.6/}
        \pie[color = {blue!50, yellow!60, pink!50,orange!70,red!100, green!20}, radius=1, hide number, pos={4.4,0}, text = legend]{56/H$_{2}$O, 14.7/He, 16.4/N$_{2}$, 10/CO$_{2}$, 11/CH$_{4}$, 1.9/H$_2$}
    \end{tikzpicture}
    \caption{Gas mixtures by mixing ratio from \citet{Khare1984OpticalProp} at room temperature (left) and \citet{He2023Methods} (middle and right) at 300\,K and 400\,K respectively.}
    \end{subfigure}
     
    \begin{subfigure}[b]{\columnwidth}
        \begin{tikzpicture}
        \pie[color = {pink!50,red!100}, radius=1, hide number,pos={0,0}]{95/, 5/}
        \pie[color = {pink!50,orange!70, red!100}, radius=1, hide number,pos={2.2,0}]{90/, 5/, 5/}
        \pie[color = {pink!50,orange!70, red!100}, radius=1,pos={4.4,0}, hide number, text=legend]{90/N$_{2}$, 8/CO$_{2}$, 2/CH$_{4}$};
        \end{tikzpicture}
        \caption{Gas mixtures by mixing ratio from \citet{Gavilan2018Haze}/\citet{Corrales2023HazesCO} with C/O ratio = $\infty$ (left), C/O = 1 (middle), and C/O = 0.625 (right) respectively}
    \end{subfigure}
\caption{(a) The gas mixtures in this work represent approximately equilibrium compositions at 1000x solar metallicity, scaled for experimental simplicity and generalized atmospheres. The two haze samples differ in the minor carbon source, as bolded. (b) Gas mixtures from \citet{Khare1984OpticalProp} (left) of a Titan-like atmosphere, and \citet{He2023Methods} (middle and right) of 1000x solar metallicity exoplanet atmospheres, including inert gasses. (c) Gas mixtures created in \citet{Gavilan2018Haze} used to make tholins grown in environments with C/O ratios adjusted to values that may be seen in exoplanet atmospheres. Real refractive indices for these tholins are calculated in \citet{Corrales2023HazesCO} and imaginary refractive indices in \citet{Gavilan2018Haze}.}
\label{fig:GasPieCharts}
\end{figure}
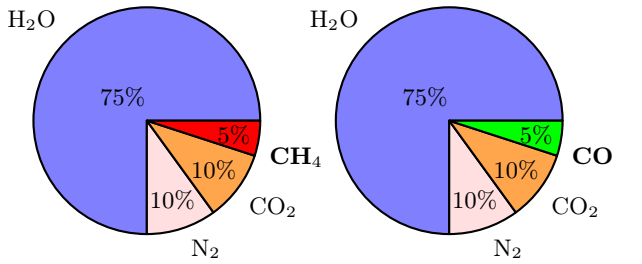
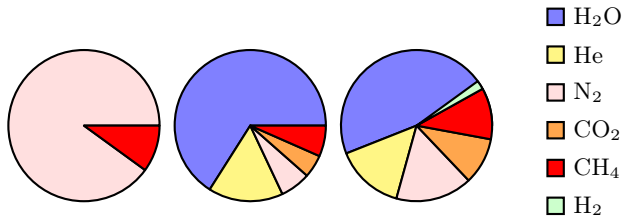
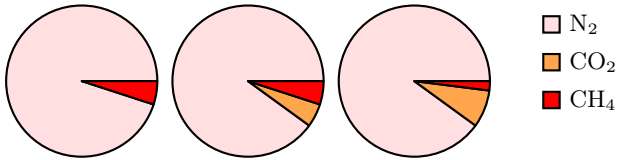

Since detailed compositions of exoplanet atmospheres are not well constrained, we rely on chemical equilibrium calculations to guide the starting atmospheric composition \cite[]{Moses2013Chemistry}. 
The laboratory hazes in this work were made to be representative of a water world like atmosphere, informed by 1000$\times$ solar metallicity abundances, as motivated by previous laboratory experiments \cite[]{He2018HazeRough,He2018MiniNeptune,Horst2018PHAZER}. 
The mixing ratios for our simulated atmosphere can be found in Figure \ref{fig:GasPieCharts_Huseby2025}. 

A detailed summary of the experimental setup can be found in \citet{Huseby2025Hazes} and \citet{He2017PHAZER} with a brief summary here. 
All gases in Figure ~\ref{fig:GasPieCharts_Huseby2025}, except for water vapor, were premixed into a cylinder with high-purity gases (Airgas, N$_{2}$ - 99.9997\%, CO$_{2}$ - 99.999\%, CH$_{4}$ - 99.999\%, CO - 99.999\%). This premixed composition is flowed into the PHAZER chamber (JHU, Baltimore, MD) at a rate of 2.5 standard cubic centimeters per minute (sccm). 
Water vapor (HPLC water, Fisher Chemical) is introduced into the PHAZER system at a pressure of 500Pa from water ice at 270\,K. 
The water vapor abundance is controlled by maintaining a liquid water cylinder at a specific temperature, maintained by a slush bath and monitored with a thermocouple, to achieve a desired vapor pressure. The water vapor pressure corresponds to a flow rate that maintains the desired mixing ratio when mixed with the rest of the gases.
Together, the mixture was exposed to AC glow discharge. 
Newly-formed solid particles settled in the chamber onto optical-grade MgF$_2$ substrates (diameter: 25 mm, thickness: 1 mm, Crystran) as thin films. 
After 72 hours, the gas flow and glow discharge was turned off, and samples were kept under vacuum for 48 hours to remove any potential volatile components. 
The haze products were kept in an N$_{2}$-purged glove box in sealed containers under aluminum foil wrapping to avoid any contamination from light sources and Earth's atmosphere. 
The haze sample films were transferred to plastic cases, which were sealed with parafilm and covered with aluminum foil for storage.
The samples were transported, still sealed and covered, to the University of Arizona for the UV irradiation experiments. 
Throughout the rest of this work, we refer to the solid sample that was generated under the conditions of the left panel of Figure \ref{fig:GasPieCharts_Huseby2025} as "the 5\% CH$_4$-derived haze sample" and the sample generated under the conditions of the right panel of Figure \ref{fig:GasPieCharts_Huseby2025} as "the 5\% CO-derived haze sample". 

\subsection{FTIR Spectroscopy Measurements and Calculations}\label{subsec:FTIR}
\begin{figure*}[ht]
\centering
\includegraphics[width=\textwidth]{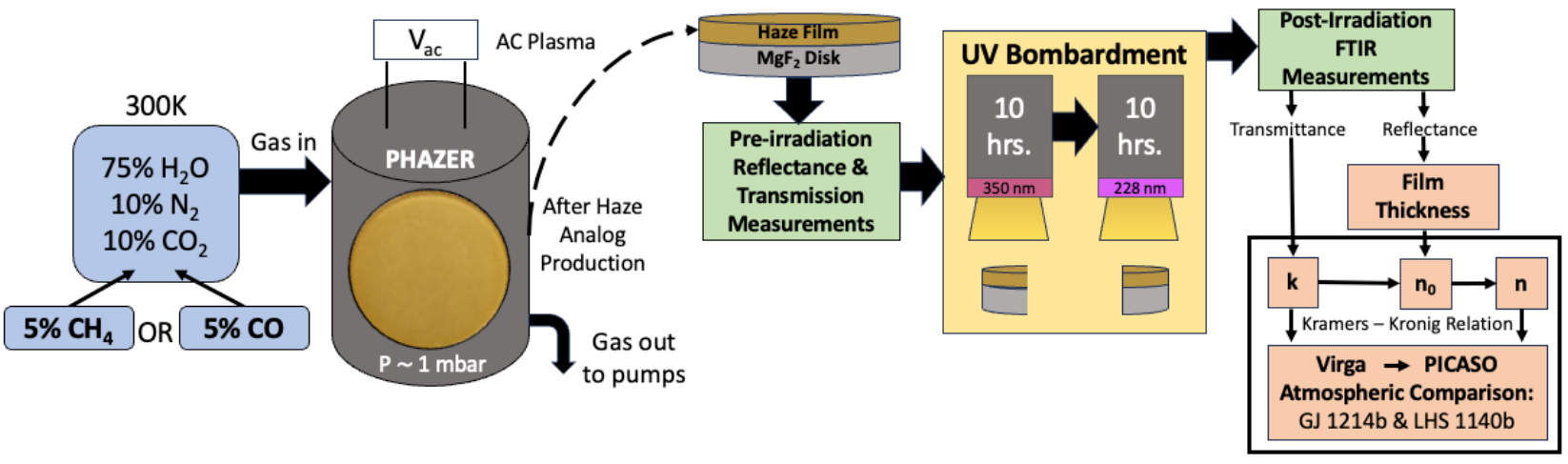}
\caption{Cartoon schematic of the experimental setup (See methods and \citealt{Huseby2025Hazes}), simulated atmospheric compositions and conditions, UV bombardment process, measurements, and experimental outcomes. We use the spectral results from our previous work to determine the film thickness of the samples. In this work, highlighted in the black box, the real ($n$) and imaginary ($k$) refractive indices are derived using the Kramers - Kronig relation between $n$ and $k$. After deriving the optical constants, we use Virga to calculate mie scattering coefficients, and generate transmission spectra using PICASO to observe the changes to exoplanets GJ 1214b and LHS 1140b modeled spectra pre- and post-irradiation.}
\label{fig:ProductionSchematic}
\end{figure*} 

\subsubsection{FTIR Measurements and UV Bombardment}
For irradiation and measurement, the substrates are placed in a room-temperature (294K)  N$_2$-purged sample chamber (99.999\%, AirProducts) in the Fourier Transform Infrared (FTIR) spectrometer (Nicolet iS 50R Benchtop FTIR, University of Arizona, Tucson, AZ) on a sliding stage and custom sample holder that contained two standards and the sample.
The sliding stage and custom sample holder enabled standards and samples to be moved in and out of measurement and UV irradiation positions of both the spectrometer and light source without breaking N$_2$ purge and to better focus the UV light onto the samples. 
The N$_2$ purge ensures no water or molecular oxygen are present during the experiment. 
Therefore any chemical changes we observe are due to the UV-irradiation and to the haze itself. 
Before irradiation, we took visible to thermal infrared  24,000 to 1250 cm$^{-1}$ (0.5 - 8 $\mu$m) reflectance and transmission spectra with the FTIR spectrometer. 
For a list of instrument configurations, see \citet{Huseby2025Hazes}. 

We irradiated the samples with two different bandpasses of light (215 - 245 nm, peak: 228 nm, 20\% throughput, Andover Corporation; and 320 - 380 nm, peak: 350 nm, 80\% throughput, Baader Planetarium) for 10 hrs per filter. 
The filters had different peak wavelengths and throughputs to compare a representative higher and lower energy flare impacting the atmosphere of the planet. 
Further discussion of the comparison between flaring energies and our experimental output and duration is described in detail in \citet{Huseby2025Hazes}. 
One half of the sample was bombarded with UV light through the 350 nm filter, where after irradiation the sample was rotated, and bombarded through the 228 nm filter. 
Transmission and reflectance measurements were taken after each irradiation cycle.
We compared all measurements against a blank MgF$_2$ plate to remove any background noise during the irradiation process and effects from the substrate.

\subsubsection{Optical Constants Calculations} \label{subsec:Opticalcalculation}
In order to calculate optical constants \textit{n} and \textit{k} from the transmission spectra ($T$) and reflectance spectra ($R$) of the film, we obtain film thicknesses for our samples pre- and post-irradiation following the interference fringing procedure of a thin film from \citet{1987NeriInterference} in our reflectance measurements \citep{Huseby2025Hazes}.
Before irradiation, we obtain a thickness of 0.911 $\pm$ 0.094 ${\mu}$m for the 5\% CH$_4$-derived haze sample, and a thickness of 0.368 $\pm$ 0.037 ${\mu}$m for the 5\% CO-derived haze sample.
The 5\% CO-derived haze sample is thinner than the 5\% CH$_4$-derived haze sample due to a lower haze production rate with CO compared to CH$_4$, as is common for such compositions \citep{He2017PHAZER,He2018HazeRough,Moran2022Triton}. 
After irradiation, we calculate a thickness of 0.862 $\pm$ 0.021 ${\mu}$m for the 5\% CH$_4$-derived haze sample across the 350 nm filter and a thickness of 0.891 $\pm$ 0.056 ${\mu}$m for the 5\% CH$_4$-derived haze sample across the 228 nm filter. 
We see no changes in the thickness of the 5\% CO-derived haze sample within the precision of our measurements. 
For further discussion, see \citet{Huseby2025Hazes}. 

We calculate $k$ given our transmittance, reflectance, and thickness results using Equation \ref{eq:extcoe} from \citet{Khare1984OpticalProp,He2022Titan}:

\begin{align}
    \label{eq:extcoe} 
    k = \frac{\lambda}{4 \pi t} \ln \frac{1-R}{T}
\end{align} 

\noindent where $\lambda$ is wavelength, $t$ is the thickness of the sample, $R$ is the reflectance of the sample, and $T$ is the transmittance of the sample. 

In strong light scattering scenarios, the resulting calculations of optical constants could lead to an underestimation of the true optical constants \citep{Ramirez2002reflections}. 
The surfaces of our thin films are smooth, with a post-irradiation calculated rms surface roughness of 17.67 nm, or 4.6\% of the total film thickness, for the 5\% CO-derived haze sample and 3.08 nm, or 0.3\% of the total film thickness, for the 5\% CH$_4$-derived haze sample, smoother because more sample was present \citep{Huseby2025Hazes}. 
Therefore we neglect the potential scattering loss for our experimental samples, as shown in \citet{He2022Titan}. 

We use the open-source numerical solver \textit{PyElli} \citep{muller2025Pyelli} to solve the Kramers-Kronig relations using the Maclaurin's Formula \citep{Maclaurin1988formula}, as seen in Equation \ref{eq:maclauren}. 
We use our derived $k$ values from 24000 to 1250 cm$^{-1}$ (0.5 to 8 $\mu m$) in the integrand. Kramers-Kronig using Maclaurin's formula is given by:

\begin{align}
    \label{eq:maclauren} 
    \Delta n(\lambda) = n(\lambda_i) - n(\infty) =  n_0 + \frac{2}{\pi} \int_0^{\infty} \frac{\lambda \, k(\lambda)}{1 - \frac{\lambda^2}{\lambda_0^2}} \, d\lambda
\end{align}

\noindent Where $n$ is the real part of the refractive indices, $k$ is the imaginary part of the refractive indices, and $\lambda$ is our wavelength range. 
We use an anchor point, $n_0$, which is the n value of our sample, and which is required to obtain an analytical solution using Kramers-Kronig \citep{Khare1984OpticalProp,He2022Titan}. It is a necessary parameter in order to scale n over our spectral range: 

\begin{align}
    \label{eq:realref}
    n_0 = \frac{1 + R + \sqrt{4R - k^2 (1 - R^2)}}{1 - R}
\end{align}

\begin{figure*}[ht!]
    \centering
    \includegraphics[width=0.9\textwidth]{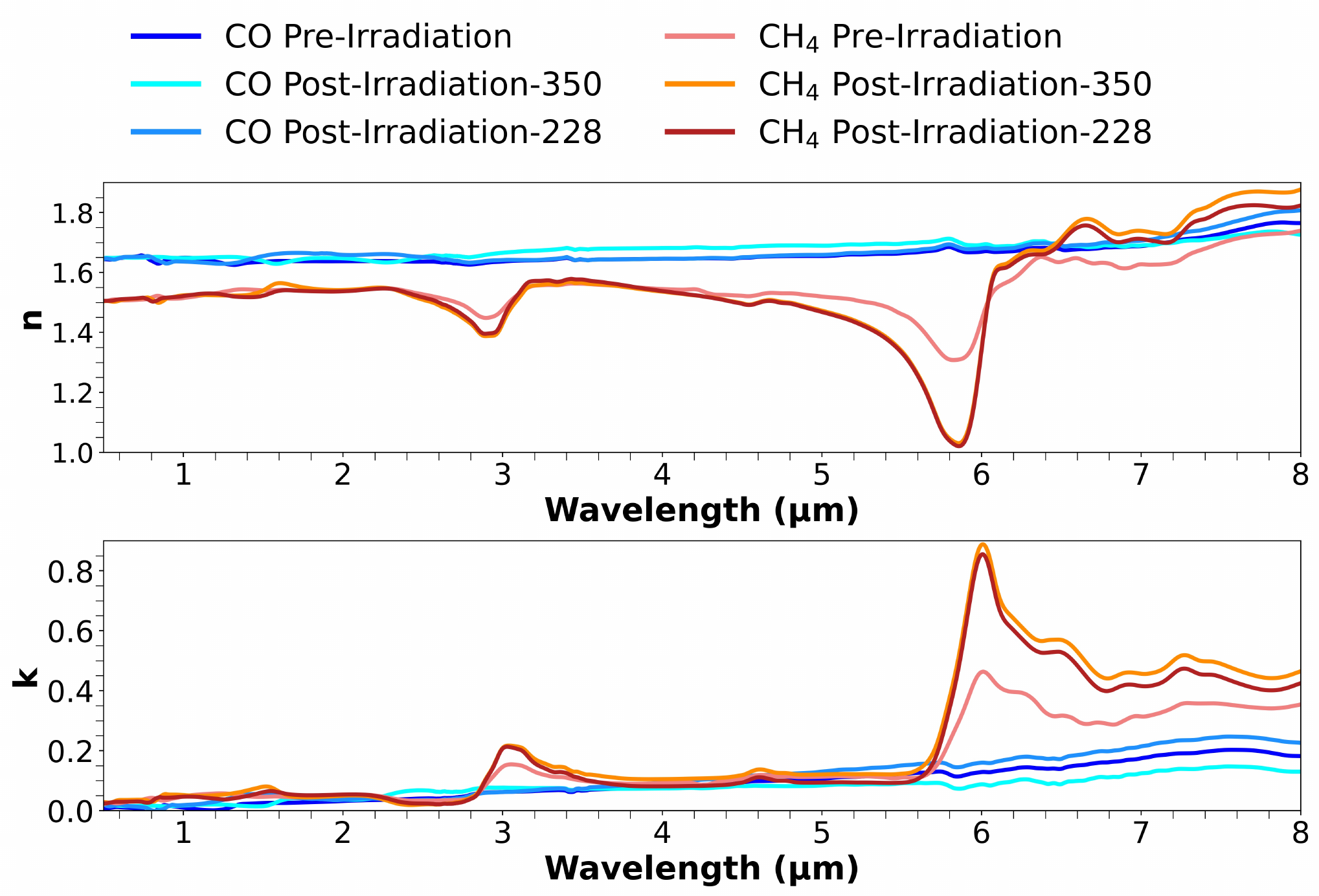}
    \caption{Plot of optical constants of both our 5\% CO and 5\% CH$_4$-derived haze samples in the visible to mid-IR wavelength region (0.5 to 8 $\mu$m, 24000 to 1250 cm$^{-1}$). Numbers 228 and 350 corresponds to the peak irradiation wavelength in nm simulating the flare. Top: Real ($n$) part of the refractive indices and Bottom: imaginary ($k$) part of the refractive indices. There are large differences between both the samples and pre- and post-irradiation.}
    \label{fig:Comparativetotal}
\end{figure*}

The anchor point is determined using the reflectance and extinction coefficients in Equation \ref{eq:realref} at a chosen wavelength of 2.9128 $\mu$m \citep{Toon1994Uncertainty,He2022Titan}. 
We find the $n_0$ anchor point for the 5\% CH$_4$-derived haze sample to be 1.362 and an anchor point value of 1.498 for the 5\% CO-derived haze sample.
We plot the $k$ and $n$ values of our two samples pre- and post-irradiation in Figure \ref{fig:Comparativetotal}. We recognize there will be potential error in the haze sample $n$ and $k$ values including error from film thickness and scattering between the film and substrate. We point the reader to the appendix for further discussion and how we account for those errors.





\subsection{Atmospheric Modeling} 
\label{subsec:Atmo model}
We use the open source radiative transfer code PICASO (Version 3.0; \citealt{2022BatalhaPICASO,Mukherjee2023PICASO}) and aerosol modeling code Virga (Version 1.0; \citealt{Batalha2025VIRGAOFFICIAL}) to compute Mie scattering coefficients from our optical constants and implement them into a resulting transmission spectrum model.
In this work we model two planets, GJ 1214b and LHS 1140b, for the impact of our calculated haze optical properties to relevant atmospheric transmission spectra (See appendix for all input parameter values). 

We use the package PyMieScatt \cite[]{Sumlin2018PyMieScatt} called within Virga (Version 0.1) to generate wavelength-dependent Mie scattering coefficients. 
We input custom haze particle size distributions using those measured in a 1000x solar metallicity 300\,K haze experiment \citep{He2018MiniNeptune}, as implemented in \citet{He2023Methods}.  
We note that using Mie scattering is an approximation, as this technique uses spherical particles rather than the different aerosol particle morphologies that are likely found in actual exoplanet atmospheres \citep{He2018MiniNeptune,Adams2019Aggregates,Ohno2020aggregate,Moran2025Aggregates}. 

In addition to the $n$ and $k$ values and haze particle size distribution, inputs for the vertical extent of the haze layer within the atmosphere and mass loading of the haze layer are necessary to compute a full atmospheric transmission spectrum. 
However, Virga currently does not have the capability to self-consistently compute photochemical haze profiles.
Therefore, we follow methods for modeling the atmospheres of Titan and super-Earth and sub-Neptune exoplanet modeling, inputting a homogeneous haze layer in PICASO from 0.1 bar to 0.1 $\mu$bar in pressure \cite[]{Kawashima2018Hazes,Kawashima2019Hazes,He2023Methods} with a haze mass loading of $\sim$25 particles cm$^{-3}$ \citep{Lavvas2009TitanMeso,He2023Methods}, for a final column density of 4.5 x 10$^{11}$ particles cm$^{-2}$. While a more realistic haze density profile would likely vary with altitude \citep{Lavvas2019GJ1214}, such considerations are beyond the scope of our optical property comparisons. 
Other inputs include stellar radius, stellar metallicity, stellar effective temperature (T$_{\rm eff}$), and stellar gravity; planetary mass and radius; an isothermal pressure-temperature profile of the planet atmosphere; and an initial atmospheric gas mixture representative of our haze atmospheric composition. 
These parameters (see Appendix \ref{appendix:params} for full list) are incorporated into PICASO using an equilibrium temperature of 600\,K for GJ 1214b \cite[]{Cloutier2021GJ1214b} and 300\,K for LHS 1140b \citep{Cadieux2024lhs1140}. 
We then generate two sets of spectra using  our derived haze optical properties pre- and post-irradiation. 

\begin{figure*}[ht!]
    \centering
    \includegraphics[width=0.9\textwidth]{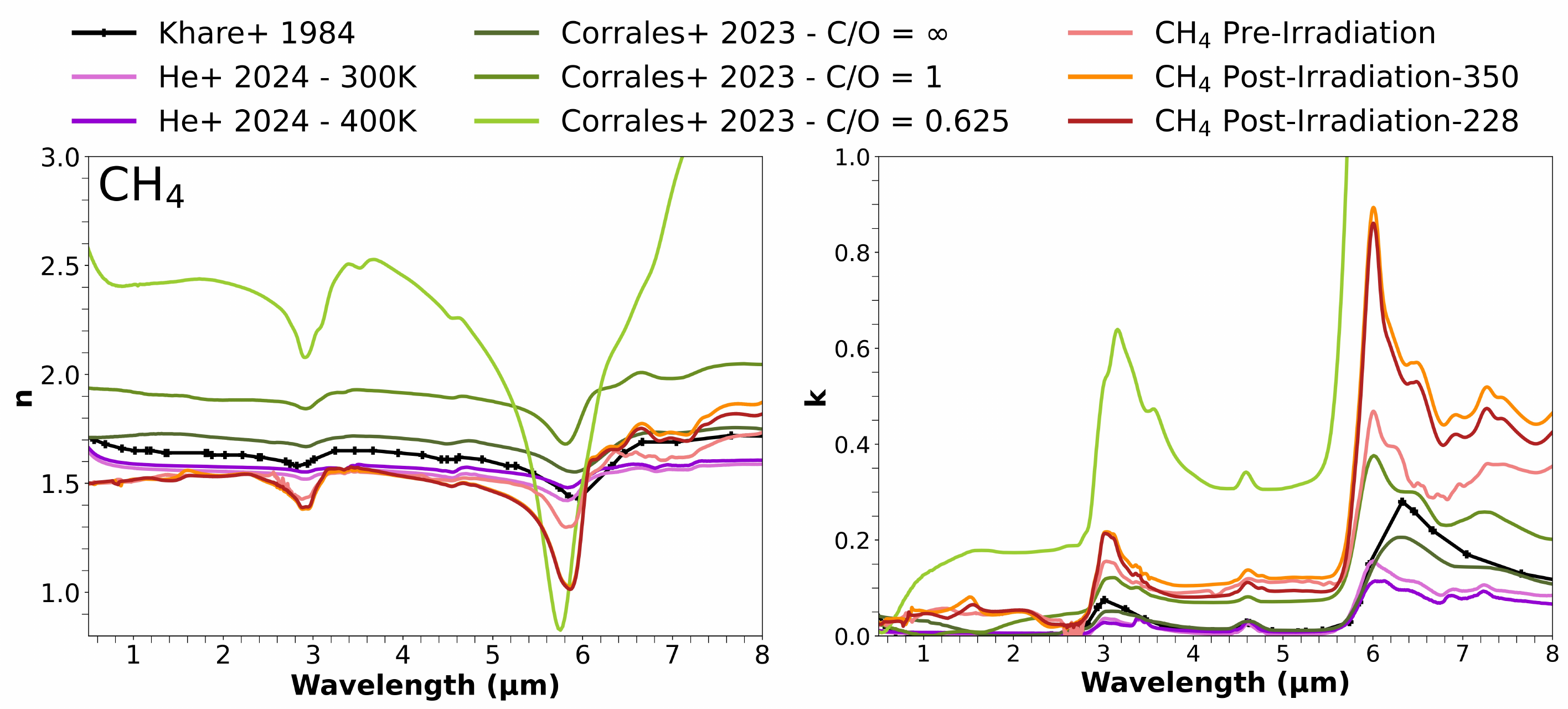}
    \includegraphics[width=0.9\textwidth]{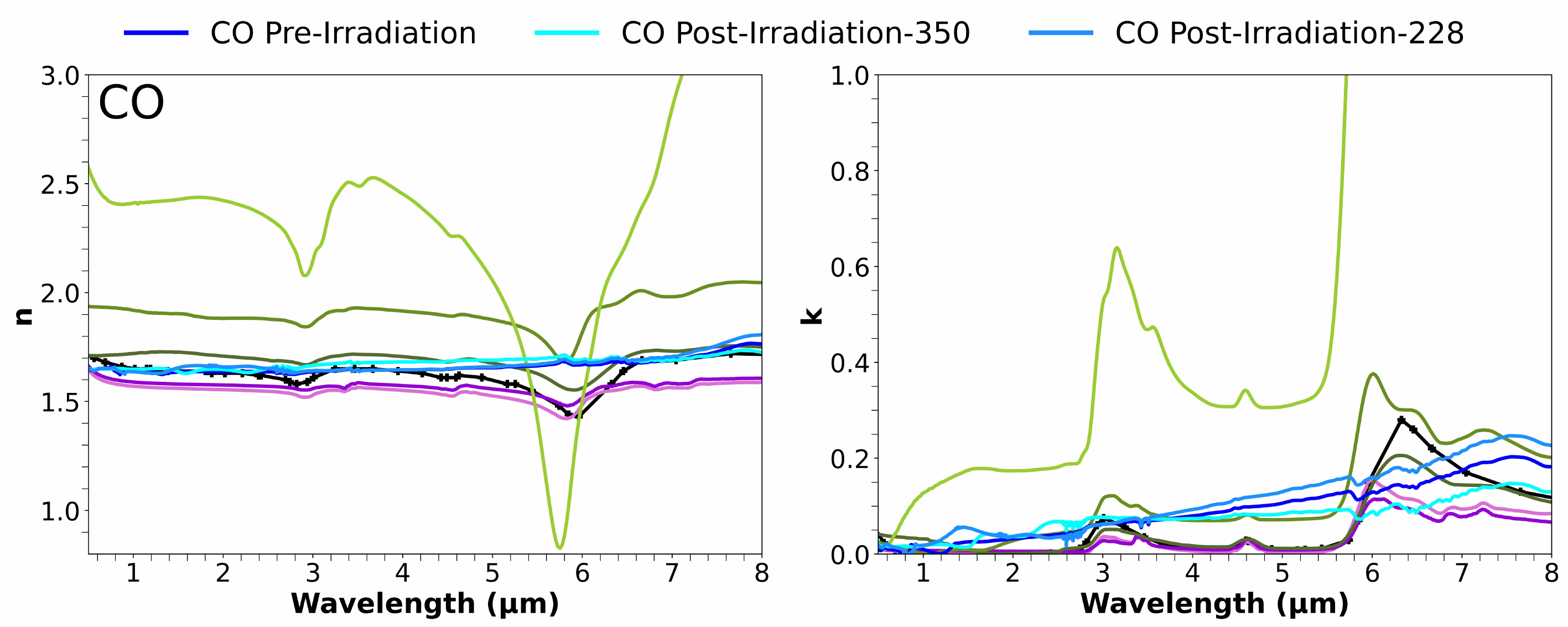}
    \caption{Comparison of optical constants of our 5\% CH$_4$ (top) and 5\% CO (bottom) atmosphere haze sample to previous studies in the visible to mid-IR wavelength region (0.5 to 8 $\mu$m, 24000 to 1250 cm$^{-1}$). The \hyperref[k84ref]{K84} sample is found in black, the \hyperref[h23ref]{H24} 300\,K sample in light purple, 400\,K sample in dark purple, and the \hyperref[g18ref]{G18/}\hyperref[c23ref]{C23} samples in varying green colors depending on C/O ratio are overplotted for reference. Left: real ($n$) part of the refractive indices and Right: imaginary ($k$) part of the refractive indices. Each of these optical properties are compositionally different, as shown in Figure \ref{fig:GasPieCharts}, and produce similarly shaped yet varying strengths of complex refractive indices.}
\label{fig:CH4/COComparativeplot}
\end{figure*}

We compare these models using all the same atmospheric and planetary parameters, but substituting the optical constants of \citet{Khare1984OpticalProp,He2023Methods} and \citet{Gavilan2018Haze}/\citet{Corrales2023HazesCO} for the hazes.
The purpose of the synthetic spectra simulations we perform are to demonstrate the effect of irradiation on the haze optical properties and the resulting atmospheric transmission spectra. 
We also perform a reduced chi-squared analysis between our simulations and the observations to quantify the influence of different optical constants and highlight the importance of experimental work. 
We note that this work is not meant to be a full parameter study, nor are we fitting the models to the observations made of these potential water-dominated planets to interpret the data. 
We made simplifying assumptions involving the pressure-temperature profile, haze particle radius and distribution, and haze mass loading.
Future modeling efforts that use self-consistent iterative pressure-temperature profiles and haze radiative effects (e.g., \citealt{Morley2015clouds,Lavvas2019GJ1214,Kempton2023Waterworlds}) in addition to aerosol microphysics (e.g., \citealt{Yoon2014Benzene,Kiefer2024CloudMorphology,Gao2018Microphysics,Gao2020Haze}) are necessary to fully describe the effects of irradiated hazes on exoplanet atmospheric observations and modeling efforts. 

\section{\textbf{Irradiated Haze Optical Constants}} \label{sec:Results}
We present newly-derived optical constants  (Figure \ref{fig:Comparativetotal}) of water-rich hazes pre- and post-irradiation over a broad wavelength range (0.5 to 8 $\mu$m, 24000 to 1250 cm$^{-1}$). We note that the $n$ values are directly derived from the $k$ values through Equation \ref{eq:maclauren}.
Comparisons of the optical constants within our CH$_4$-derived haze sample show differences in peak values pre- and post-irradiation.
The $n$ values slightly deviate from 2.5 to 3.25 $\mu$m and 4.5 to 6.25 $\mu$m, with the post-irradiation spectra values decreased. 
The $k$ values increase over the same wavelength regions.
Longward of 5.6 $\mu$m, the pre-irradiation values have a maximum of 0.47$^{+0.1}_{-0.085}$ and post-irradiation $k$ max values peak at 0.89$^{+0.1}_{-0.085}$, almost a factor of 2 larger after the irradiation period. This is different than previous work irradiating Titan-like and Earth-like hazes, which instead found decreases in absorptivity with wavelength \citep{Gavilan2018Haze,Carrasco2018VUVhaze}.
These higher $k$ values and consequently decreased $n$ values may be attributed to new complex organic bonds present in the haze sample after irradiation, including potential C-H, C=C, C=N, and N-H bonding, but the exact molecular bond is unknown.

The optical constants of our CO-derived haze sample show little to no difference across our wavelength range pre- and post-irradiation (Figure \ref{fig:Comparativetotal}), reflecting the fact that our samples were not significantly altered in transmittance or reflectance during the UV bombardment process \citep{Huseby2025Hazes}.
In the $k$ values after 5.7 $\mu$m, the pre-irradiation and post-irradiation across the 228 nm filter begin to increase at different rates, peaking at 7.5 $\mu$m. 
The largest changes occur at this peak, where we have a $k$ value of 0.15 pre-irradiation, a decreased $k$ value of 0.1 post-irradiation across the 350 nm filter, and an increased $k$ value of 0.2 post-irradiation across the 228 nm filter. 
This peak is largely due to complex organic oxygen absorption, mainly from O-H phenol in addition to multiple C-H bending modes, where the feature changes slightly during irradiation.
Though these slight variations are present, we see no significant changes in the $n$ or $k$ values of the 5\% CO-derived haze sample given our experimental uncertainties.


Our experiments show that there are large differences between the optical constants of our two laboratory samples (Figure \ref{fig:Comparativetotal}; see also Figure \ref{fig:errorprop} in Appendix \ref{appendix:errs} showing uncertainties). 
From 0.5 to 6 $\mu$m, our 5\% CO-derived haze sample has consistently higher $n$ values than the 5\% CH$_4$-derived haze pre-irradiation values.
We see substantial $n$ and $k$ features in the 5\% CH$_4$-derived haze sample, while we mostly see flat features in the 5\% CO-derived haze sample. 

We also compare our sample optical constants with previous laboratory exoplanet haze optical constants and Titan-like haze optical constants (Figure \ref{fig:CH4/COComparativeplot}). 
This is not meant to be an exhaustive comparison, and instead we selected literature values close to our compositions and with varying oxygen content (Full compositions can be found in Figure \ref{fig:GasPieCharts}). 
Our selected literature values include \citet{Khare1984OpticalProp}\label{k84ref} (hereafter \hyperref[k84ref]{K84}), which provides optical constants of Titan-like hazes; \citet{He2023Methods}\label{h23ref} (hereafter \hyperref[h23ref]{H24}), which provide optical constants for 1000x metallicity exoplanet hazes at both 300\,K and 400\,K; and \citet{Gavilan2018Haze}\label{g18ref} $k$ values and \citet{Corrales2023HazesCO}\label{c23ref} (hereafter \hyperref[g18ref]{G18/}\hyperref[c23ref]{C23}) $n$ values, which provide optical constants of Titan-like or early Earth-like hazes with varying C/O ratios, some of which may be likely in terrestrial exoplanet atmospheres. 
We find that our experiments create unique optical constants that broadly match previous literature values, but show important differences.

The shape of the real refractive indices across the experiments are similar to the 5\% CH$_4$-derived haze sample, but vary significantly in the magnitude of molecular stretching and vibrational bands throughout the wavelength range (Figure \ref{fig:CH4/COComparativeplot}). 
Our $n$ values are generally smaller than the selected comparison samples, though the feature amplitudes are most comparable to the C/O = 0.625 in \hyperref[g18ref]{G18/}\hyperref[c23ref]{C23}, which sees these large features due to the amount of oxygen present in the sample (Figure \ref{fig:CH4/COComparativeplot}). 
We hypothesize that our sample has more oxygen rich absorbing bands post irradiation compared to many of the selected samples. 
The UV bombardment changed the 6 $\mu$m molecular band significantly in our sample, where post-irradiation values now more closely align with the lowest \hyperref[g18ref]{G18/}\hyperref[c23ref]{C23} C/O ratio sample. 
The $k$ values are similar across the experiments until 6.5 $\mu$m, where we see more structural changes (Figure \ref{fig:CH4/COComparativeplot}). 
While our haze sample is likely becoming absorbing through increased bonding with oxygen during irradiation, indicated by the $k$ values that lie between \hyperref[g18ref]{G18/}\hyperref[c23ref]{C23}'s C/O = 1 and C/O = 0.625 samples, it does not approach the strength of the latter, even post-irradiation. 
This follows previous thought that in the synthesis phase, if hazes are formed with more oxygen-rich molecules, specifically with the inclusion of CO$_2$ in the sample, the hazes are more absorbing \citep{Gavilan2018Haze,Moran2020SuperEarth,Corrales2023HazesCO}. 
The shape of the $n$ and $k$ values are significantly different for the 5\% CO-derived haze sample in comparison to previous samples (Figure \ref{fig:CH4/COComparativeplot}). 
The inclusion of CO instead of CH$_4$ in the original synthesis gives almost constant $n$ and $k$ values. Previous studies \citep{Ugelow2018oxygen,Moran2022Triton} demonstrate that the inclusion of CO creates less haze overall and the haze produced is more carbon-rich. 
This likely occurs because CO is less photochemically reactive than CH$_4$. 
None of our selected literature values have CO, explaining the differences in observed absorption features. 
More comparisons between CO and other similar work can be found in the very recent work of \citet{Drant2026setups}.

\begin{figure*}[ht!]
    \centering
    \includegraphics[width=0.9\textwidth]{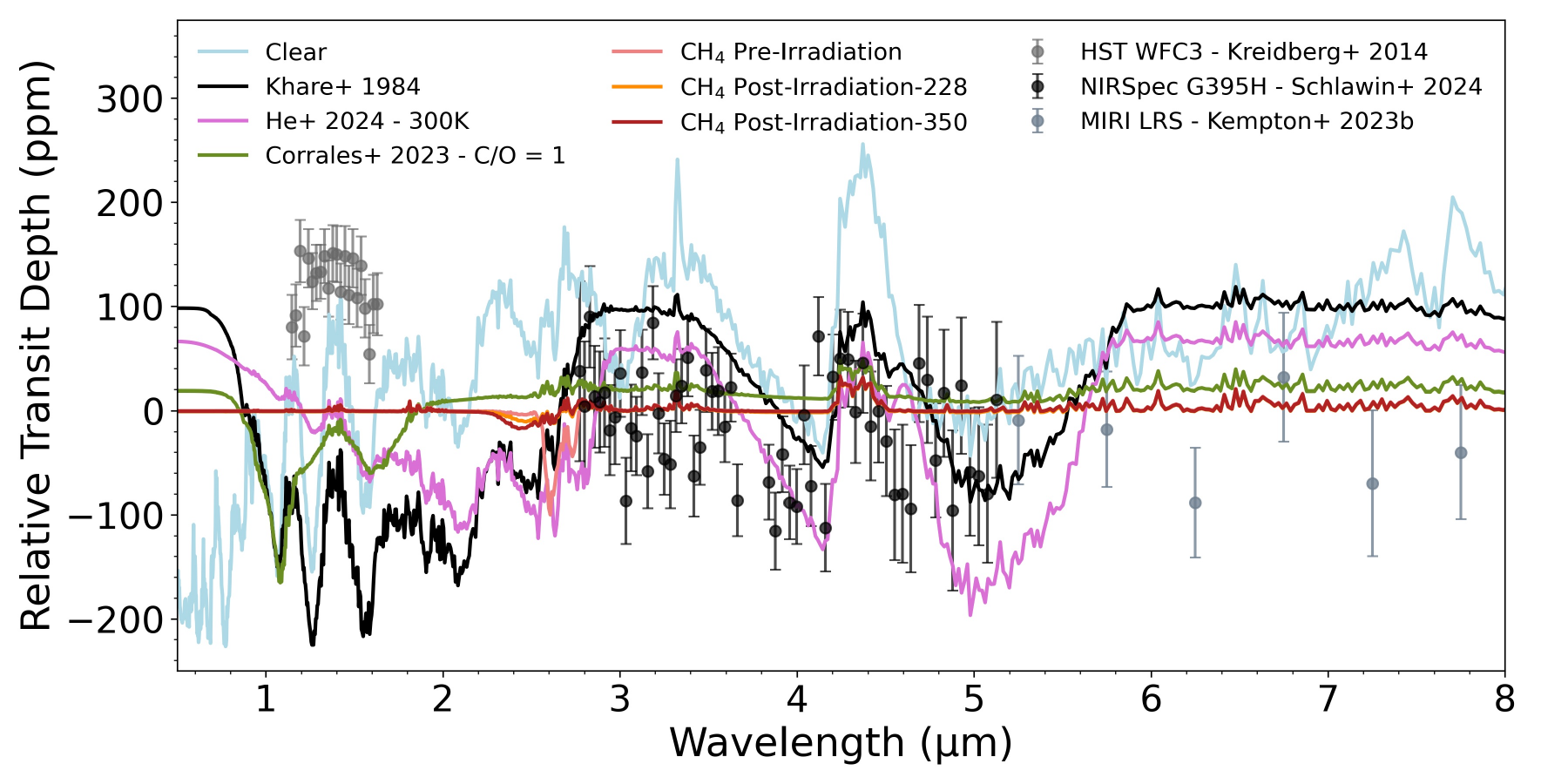}
    \includegraphics[width=0.9\textwidth]{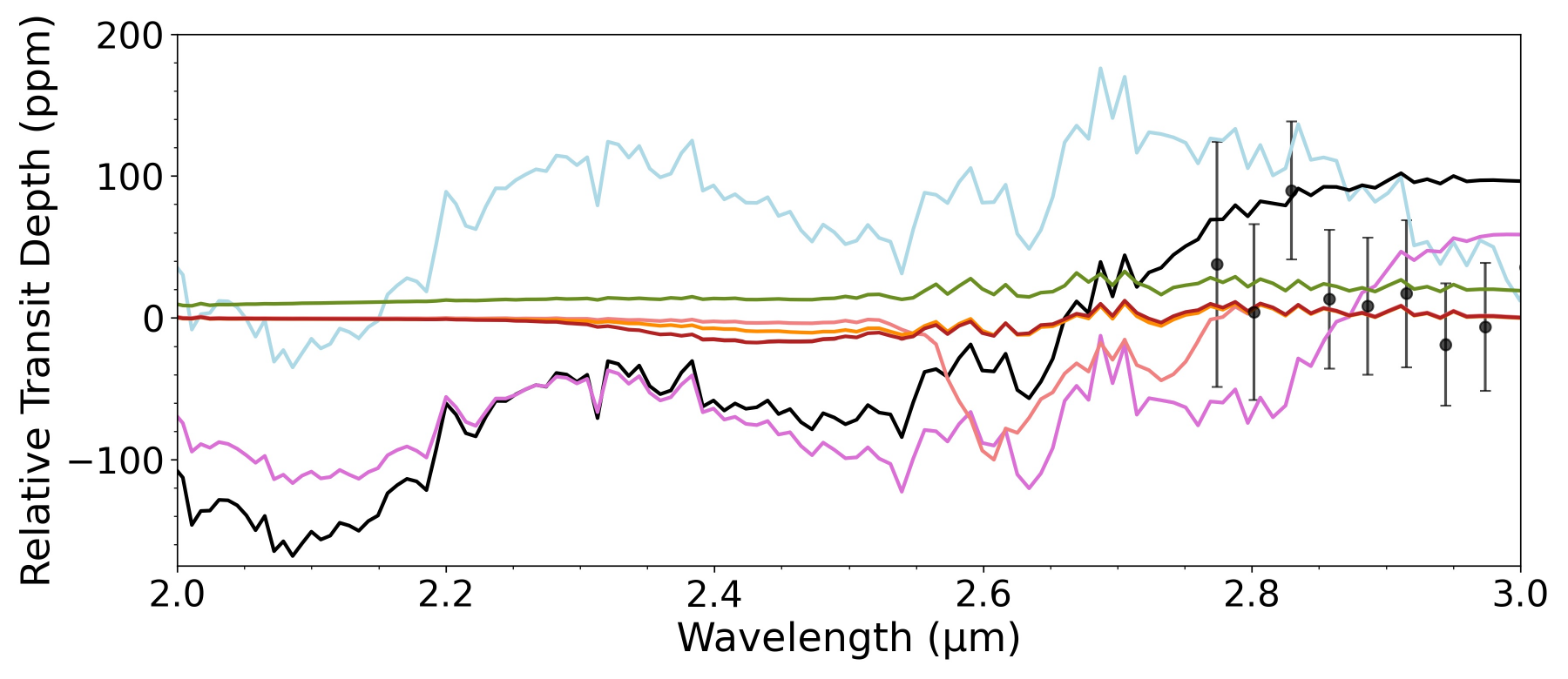}
    \caption{Transmission spectrum of GJ 1214b using the 5\% CH$_4$-derived haze sample atmospheric composition. The sample from \hyperref[k84ref]{K84}, \hyperref[g18ref]{G18/}\hyperref[c23ref]{C23} C/O = 1 sample, and \hyperref[h23ref]{H24}'s sample at 300\,K are overplotted for reference. We include the observations from \citet{Kreidberg20141214b,Kempton2023GJ1214,Schlawin2024GJ1214} with offsets for comparisons between models. The error bars account for 1$\sigma$ uncertainties in the observations. Top: Spectra from 0.5 to 8 $\mu$m, where there are large differences in the resulting transmission spectrum between the sample and previous literature optical constants. Our samples pre- and post-irradiation produce largely flat spectra. Bottom: Spectra from 2 to 3 $\mu$m. We see slight changes between our pre- and post-irradiation transmission spectra, where our hazes are letting gaseous CO$_2$ found in the simulated atmosphere to be more or less visible.}
    \label{fig:GJ1214-CH4}
\end{figure*}

The large differences between the two planet analog samples and their irradiated products show that compositionally distinct hazes provide significant changes in their optical constants. This highlights the importance of utilizing optical properties more representative to the atmosphere of study. 

\begin{figure*}[ht!]
    \centering
    \includegraphics[width=0.9\textwidth]{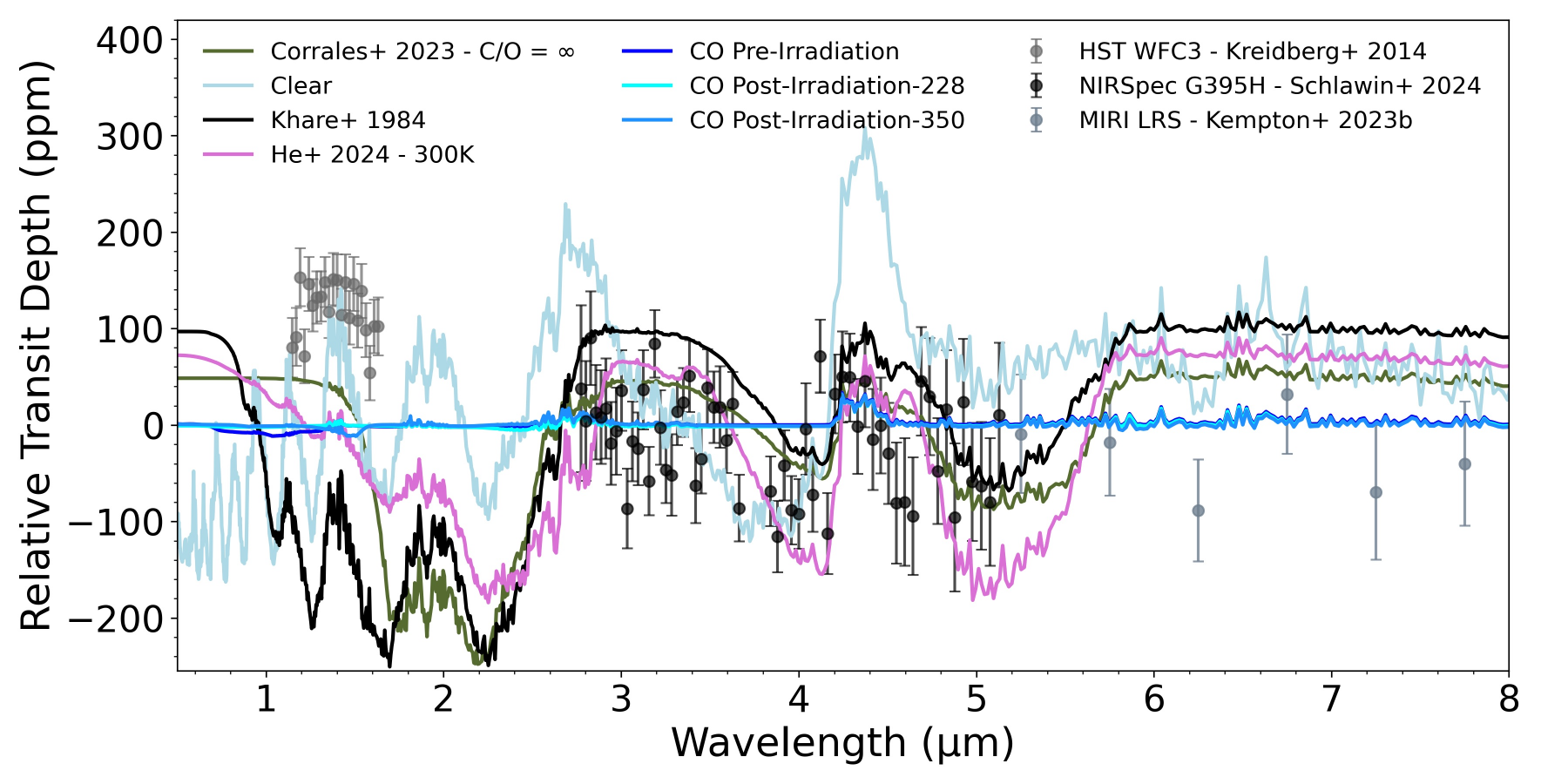}
    \caption{Transmission spectrum of GJ 1214b from 0.5 to 8 $\mu$m using the 5\% CO-derived haze sample atmospheric composition. The sample from \hyperref[k84ref]{K84}, \hyperref[g18ref]{G18/}\hyperref[c23ref]{C23}'s C/O = $\infty$ sample, and \hyperref[h23ref]{H24}'s sample at 300\,K are overplotted for reference. We include the observations from \citet{Kreidberg20141214b,Kempton2023GJ1214,Schlawin2024GJ1214} with offsets for comparisons between models. The error bars account for 1$\sigma$ uncertainties in the observations. Our samples pre- and post-irradiation make flat spectra, with slight changes between the pre- and post-irradiation across the 228 nm filter. The post-irradiation spectrum across the 350 nm filter has a consistently higher transit depth. These differences would not be detectable by existing space-based observatories like HST and JWST within the error bars.}
    \label{fig:GJ1214-CO}
\end{figure*}

\section{\textbf{Implications for Exoplanet Atmosphere Transmission}} \label{sec:TransmissionModel}
The key link between laboratory-made exoplanetary hazes and testable model predictions to observations lies in the haze optical constants, which can be directly implemented into atmospheric transmission models to compare to observations \citep[e.g.,][]{He2023Methods}.
We use the optical constants calculated here and previous literature values to explore the effects of our UV irradiated hazes on synthetic transmission spectra of potential water-world like planets. 

The aim of the analysis below is to showcase the effects that different haze optical constants have on the resulting transmission spectra, to quantify the observable changes pre- and post-irradiation, and to demonstrate the need for more representative haze optical properties for exoplanet atmospheres. 
Therefore we use set values for the haze particle distribution, number density, and pressure grid size. Statistically analyzing these spectra with observed data is outside the scope of this work, and the observations here are simply for model comparison within the error bars of current instrumentation. 

\begin{table}[]
\caption{Reduced $\chi^2$ values between models and data}
\label{tab:chisquared}

\begin{threeparttable}
\begin{tabular}{lcc}
\hline

\multicolumn{3}{c}{\textit{GJ1214b}} \\
 & 5\% CH$_4$ & 5\% CO \\
Clear Atmosphere & 12.16  & 8.84\\
Khare+ 1984     & 19.94  & 21.43\\
He+ 2024  & 6.46\tnote{1}  & 6.18\\
Corrales+ 2023   & 7.35\tnote{2}  & 4.82\tnote{3}\\
Pre-Irradiation     & 4.32 & 4.45\\
Post-Irradiation-228     & 4.30 & 4.27\\
Post-Irradiation-350    & 4.31 & 4.51\\
\hline

\multicolumn{3}{c}{\textit{LHS1140b}} \\
 & 5\% CH$_4$ & 5\% CO \\
Clear Atmosphere & 2.08  & 2.02\\
Khare+ 1984     & 2.34  & 2.42\\
He+ 2024  & 1.98\tnote{1}  & 2.00\\
Corrales+ 2023   & 2.02\tnote{2}  & 2.10\tnote{3}\\
Pre-Irradiation     & 1.92 & 1.91\\
Post-Irradiation-228     & 1.91 & 1.91\\
Post-Irradiation-350    & 1.91 & 1.91\\
\hline
\end{tabular}

\begin{tablenotes}
\item[] \hspace{2em} \tnote{1} 300K \hspace{2em}\tnote{2} C/O = 1 \hspace{2em} \tnote{3} C/O = $\infty$
\end{tablenotes}

\end{threeparttable}
\end{table}

\subsection{GJ 1214b} \label{subsec:GJ1214b}

We model the transmission spectra of the sub-Neptune planet GJ 1214b, a potentially hazy, water-dominated atmosphere as revealed by recent JWST \citep{Kempton2023GJ1214,Schlawin2024GJ1214} and significant HST observations \citep{Kreidberg20141214b}.
We compare transmission spectra with haze optical constants including the C/O~=~1 (most analogous to our CH$_4$-derived haze sample, see Figure \ref{fig:CH4/COComparativeplot}, top), and C/O = $\infty$ (most analogous to our CO-derived haze sample, see Figure \ref{fig:CH4/COComparativeplot}, bottom) samples of \hyperref[g18ref]{G18/}\hyperref[c23ref]{C23}, the 300\,K sample of \hyperref[h23ref]{H24}, and the \hyperref[k84ref]{K84} sample.
We choose the C/O~=~1 and C/O~=~$\infty$ samples from a $\chi^2$ analysis of the $n$ and $k$ values found in Figure \ref{fig:CH4/COComparativeplot} to determine the closest fitting optical constants to our sample pre- and post-irradiation.
We create a relative transit depth by subtracting the average of each set of observations and model transit depths so we have a more even comparison between models and changes are easily quantifiable. 

\begin{figure*}[ht!]
    \centering
    \includegraphics[width=0.9\textwidth]{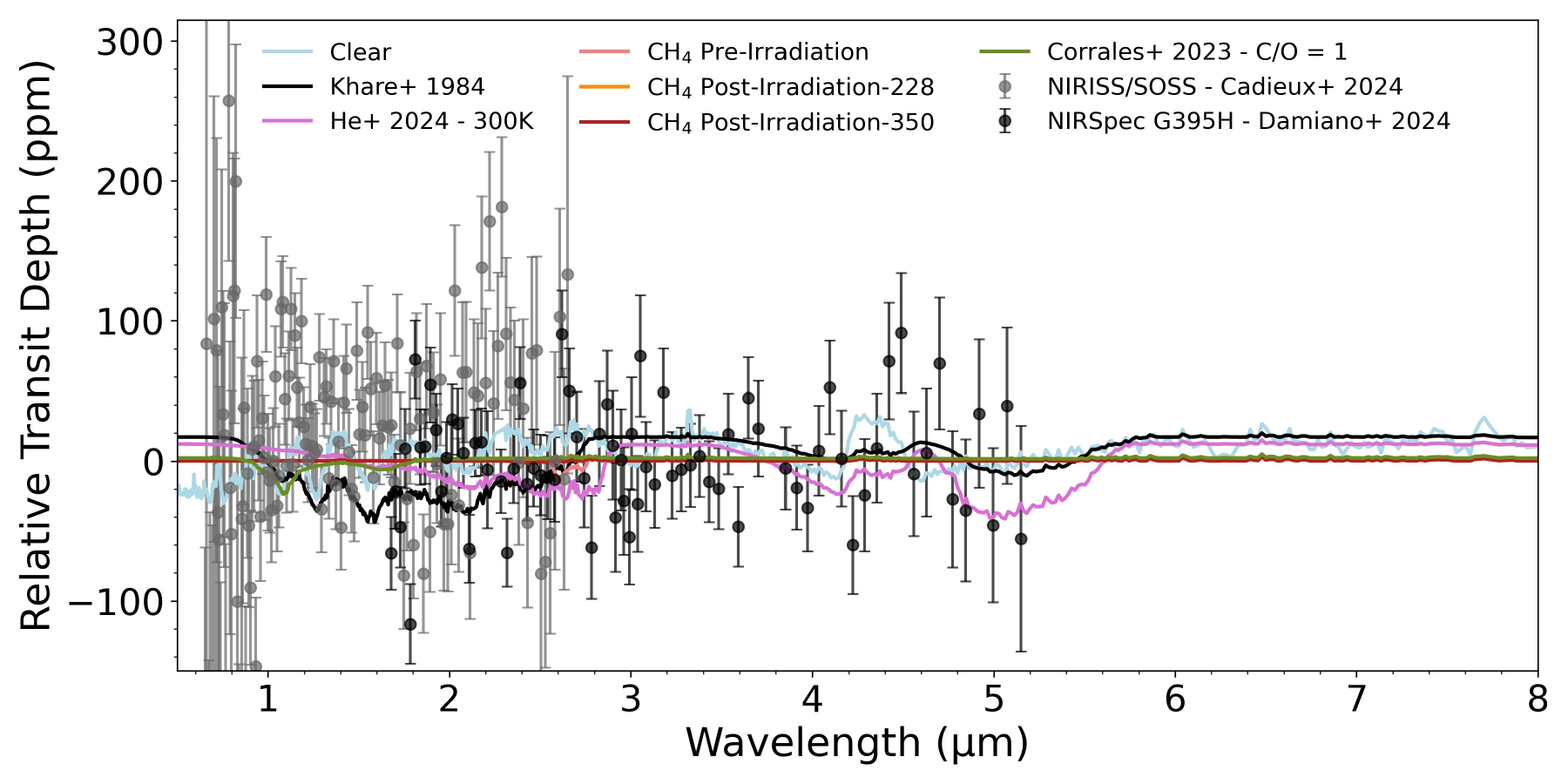}
    \includegraphics[width=0.9\textwidth]{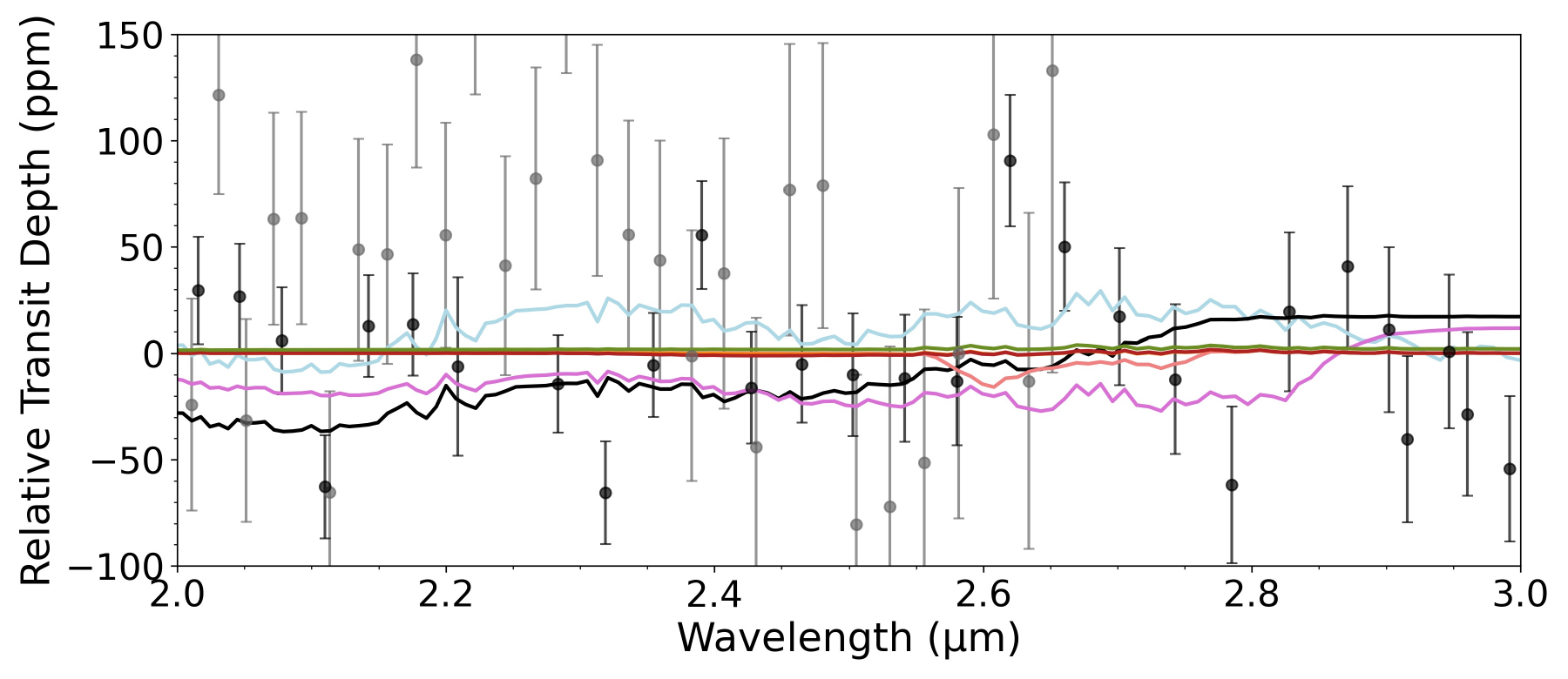}
    \caption{Transmission spectrum of LHS 1140b using the 5\% CH$_4$-derived haze sample atmospheric composition. The sample from \hyperref[k84ref]{K84}, \hyperref[g18ref]{G18/}\hyperref[c23ref]{C23}'s C/O = 1 sample, and \hyperref[h23ref]{H24}'s sample at 300\,K are overplotted for reference. We include the observations from \citet{Cadieux2024lhs1140,Damiano2024LHS1140b} with offsets for comparisons between models. The error bars account for 1$\sigma$ uncertainties in the observations. Top: Spectra from 0.5 to 8 $\mu$m, where there are differences in the resulting transmission spectrum between optical constants, and slight differences between our sample pre- and post-irradiation. Bottom: Spectra from 2 to 3 $\mu$m, where we see slight changes between our pre- and post-irradiation transmission spectra. These differences are not detectable by existing space-based observatories like HST and JWST due to the planet's small scale height.}
    \label{fig:LHS1140b-CH4}
\end{figure*}

Our simulated spectra of GJ 1214b show that all haze cases mutes the transmission spectra, but there are differing effects and magnitudes of muting depending on the specific optical properties chosen for the haze (Figures \ref{fig:GJ1214-CH4} and \ref{fig:GJ1214-CO}). 
Our samples largely create flat spectra and are more absorbing than other organic hazes found in the literature, creating a larger muting effect. 

Comparing to observations, we compute a reduced $\chi^2$ to compare the precisions needed to distinguish optical properties rather than interpret which model is best for a particular planet (Table \ref{tab:chisquared}). 
We would not be able to distinguish pre- and post-irradiation in the HST range with the approximate precision of $\sim$30 ppm \citep{Kreidberg20141214b}. 
However, we would be able to distinguish between the clear atmosphere and the hazy atmosphere models. 
In the JWST NIRSpec G395 range, most of the scenarios match the amplitude of the features data within the error bars. 
With a precision of $\sim$18-27 ppm \citep{Schlawin2024GJ1214}, we would not be able to distinguish changes pre- and post-irradiation.
In the JWST MIRI range, all of the samples would be indistinguishable due to the large error bars and lack of spectral features present in every haze sample at the modeled number density. 

In the wavelength range of 2 to 3 $\mu$m, we observe spectral features of H$_2$O and CO$_2$ from the atmosphere and amines (N-H) found in our haze sample (Figure \ref{fig:GJ1214-CH4}, bottom). 
At 2.6 $\mu$m we see a large deviation of $\sim$100 ppm between the pre- and post-irradiation values. 
This is due to the increase in absorptivity post-irradiation in the $k$ values (Figure \ref{fig:Comparativetotal}), causing a flatter spectrum. 
Observations between 2 and 3 $\mu$m, which could be completed with JWST NIRISS/SOSS, NIRCam 322W2, or NIRSpec G235, would need $\sim$20 ppm precision to be able to distinguish between a pre-irradiation and post-irradiation haze in this wavelength range. 

In our spectra of GJ1214b using the 5\% CO-derived haze sample, we find a largely flat spectrum (Figure \ref{fig:GJ1214-CO}). 
We note that the thinner film and therefore lower haze production rate for this atmospheric composition would warrant a lower number density used in the model. 
However, we keep the number density in this work consistent between the samples for direct comparison in transmission. 
The transmission spectrum post-irradiation across the 350 nm filter is higher by $\sim$15 ppm, as the $k$ values were slightly higher over most of the wavelength range (Figure \ref{fig:Comparativetotal}).
Only slight differences are seen short of 2 $\mu$m. 
All the newer experimental haze optical constant atmospheric models fit the existing data better than the models of a clear atmosphere or with hazes using \citet{Khare1984OpticalProp} optical constants. The post-irradiation CH4-derived haze sample across the 228 nm filter has the lowest reduced $\chi^2$ (Table \ref{tab:chisquared}) of all models tested. However, even this lowest reduced $\chi^2$ does not match the spectral features found in the HST or JWST NIRSpec data well, due to the higher absorption of our optical properties which fully flattens the spectra beyond that of the data. 
In addition, even though the main changes in our $k$ values occur longward of 5.6 $\mu$m, they would not be resolved observationally due to the large error bars in current MIRI observations. 

\begin{figure*}[ht!]
    \centering
    \includegraphics[width=0.9\textwidth]{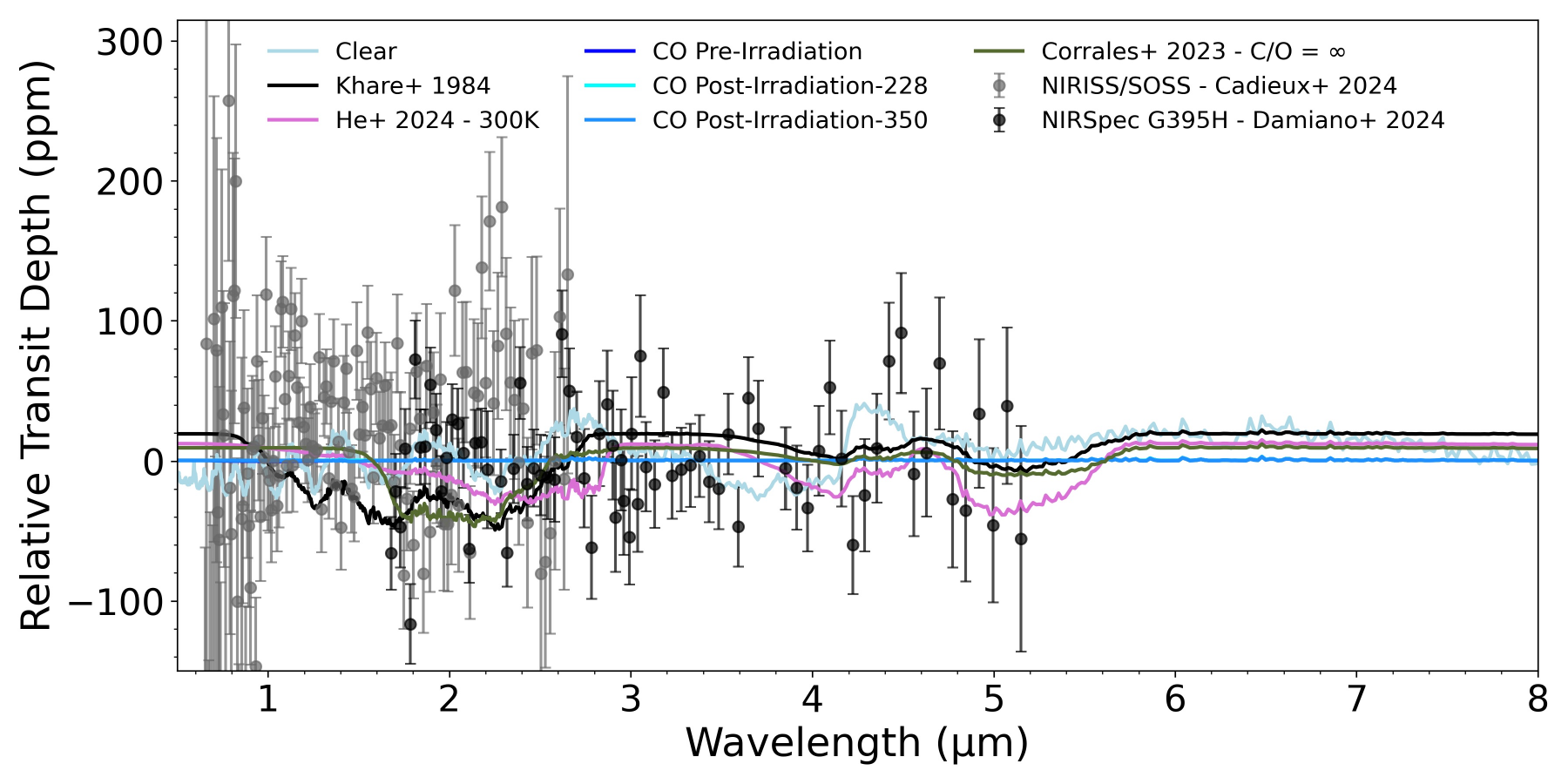}
    \caption{Transmission spectrum of LHS 1140b from 0.5 to 8 $\mu$m using the 5\% CO-derived haze sample atmospheric composition. The sample from \hyperref[k84ref]{K84}, \hyperref[g18ref]{G18/}\hyperref[c23ref]{C23}'s C/O = $\infty$ sample, and \hyperref[h23ref]{H24}'s sample at 300\,K are overplotted for reference. We include the observations from \citet{Cadieux2024lhs1140,Damiano2024LHS1140b} with offsets for comparisons between models. The error bars account for 1$\sigma$ uncertainties in the observations. Our sample pre- and post-irradiation make largely flat spectra. These differences would not be detectable by existing space-based observatories like HST and JWST within the error bars.}
    \label{fig:LHS1140b-CO}
\end{figure*}

In modeling GJ 1214b, our samples pre- and post-irradiation create largely flat spectra across our experimental wavelength range (Figure \ref{fig:GJ1214-CO}). 
There is one spectral feature that changed pre- and post-irradiation (N-H, 2.6 $\mu$m), and this change would be observable with current space-based instruments based on the change in relative transit depth and could be used as a potential marker of aging in hazes.
More broadly, small alterations in optical constants of hazes can create observable changes in a modeled transmission spectrum, and pairing optical constants most relevant to the observed planetary atmosphere has the potential to create even more accurate modeled spectra of real observations.

\subsection{LHS 1140b} \label{subsec:LHS1140b}

We also model the transmission spectra of the sub-Neptune/Super-Earth LHS 1140b. 
We chose this planet due to the fact that its radius is in the radius gap \citep{Fulton2017Gap}, but its density suggests a potential water world. It also was observed recently with JWST \citep{Cadieux2024lhs1140,Damiano2024LHS1140b} and has an equilibrium temperature similar to our experimental setup. 

Our simulated transmission spectra using the 5\% CH$_4$-derived haze sample is largely a flat line across our wavelength range (Figure \ref{fig:LHS1140b-CH4}).  
Overall, the features are more muted because of the smaller scale height of the planet atmosphere in comparison to GJ 1214b, with all reduced $\chi^2$ values close enough to each other as to be essentially statistically indistinguishable.
We see our largest differences amongst the samples between 1 and 3 $\mu$m, with changes close to $\sim$50 ppm specifically between our sample and the \hyperref[k84ref]{K84} sample. 
The optical properties of our sample are more absorbing than the other optical properties, again muting more spectral features given the same atmospheric prescription.
The closest optical properties to our transmission spectra is the \hyperref[h23ref]{H24} sample, which varies from our sample by $\sim$10 ppm and has a reduced $\chi^2$ value of 1.98 in comparison to our $\chi^2$ values of 1.92 and 1.91 pre- and post-irradiation respectively.
In comparison to observations, the large error bars ($\sim$50-150 ppm, \citealt{Cadieux2024lhs1140}) in both the NIRISS/SOSS and NIRSpec ($\sim$30 ppm, \citealt{Damiano2024LHS1140b}) data produce similar reduced $\chi^2$ values which make it difficult to distinguish which of our modeled transmission spectra are a better fit.

Within the 2 to 3 $\mu$m range, our spectra using the 5\% CH$_4$-derived haze sample shows only a minor unobservable increase in the relative transit depth pre- and post-irradiation (Figure \ref{fig:LHS1140b-CH4}, bottom). 
At 2.6 $\mu$m we see a deviation of $\sim$10 ppm between the pre- and post-irradiation values, which is much smaller than the differences found in GJ 1214b.
It is uncertain that this change would be observable even with many stacked transits of this world, and it is not currently observable within the errors of the present data as seen with the $\chi^2$ values shown in Table \ref{tab:chisquared}. 

Our simulated transmission spectra using the 5\% CO-derived haze sample shows a flat spectrum (Figure \ref{fig:LHS1140b-CO}).
Due to the small absorption features and larger error bars on both the NIRISS/SOSS and NIRSpec data, picking apart the optical properties to determine which appears to match the data best would be unobservable with current instruments, with all reduced $\chi^2$ values being similar to each other (Table \ref{tab:chisquared}). 



\section{\textbf{Conclusions}} \label{sec:Conclusion}
This study provides optical constants for laboratory-analogs of hazes in water-dominated exoplanet atmospheres that experience stellar flaring. We also provide the resulting hazy transmission spectra for two different proposed potential water-dominated planets orbiting active M-dwarf stars. When interpreting observations of exoplanets, using non-representative optical properties may lead to misinterpretations, changing the outcomes of atmospheric abundances of key species, temperature structure, dynamics, and climate feedback within an atmosphere.

Broadly, the differences between the two samples' optical constants show that compositionally distinct hazes will provide observable changes in the resulting transmission spectra depth, with differences in transit depth up to 10-100 ppm pre- and post-irradiation. 

The observable changes specifically in the N-H feature at 2.6 $\mu$m in the 5\% CH$_4$-derived haze sample pre- and post-irradiation, when applied to models of GJ 1214b, show that UV bombardment has the potential to alter the optical properties and resulting transmission spectra of water-world exoplanet hazes when CH$_4$ is present and planets have larger scale heights. 
These differences should be considered when observing water-world planets closely orbiting their stars.

\begin{acknowledgments}
 The authors gratefully acknowledge Michael Radke for providing insightful comments on the data. We also acknowledge the production of the exoplanet haze samples by the Johns Hopkins University PHAZER lab, supported by NASA under the XRP program (Grant 80NSSC20K0271), and support by NASA for this study under the SURP program (Grant 2023--048) between the Jet Propulsion Laboratory and the University of Arizona. This work was supported by a University of Arizona Postdoctoral Research Development Grant. S.E.M. is supported by NASA through the NASA Hubble Fellowship grant HST-HF2-51563 awarded by the Space Telescope Science Institute, which is operated by the Association of Universities for Research in Astronomy, Inc., for NASA, under contract NAS5-26555.
\end{acknowledgments}

\section*{Data Availability}
Data associated with this study are provided in the article and in ReData, The University of Arizona's Research Data Repository, from \dataset[DOI: 10.25422/azu.data.32226072]{https://doi.org/10.25422/azu.data.32226072}.

\bibliographystyle{aasjournalv7}
\bibliography{bibliography1}{}

\appendix

\label{Sec:appendix}

\section{Stellar and Planetary Parameters}
\label{appendix:params}

\begin{table}[h!]
    \centering
    \begin{threeparttable}
        \label{tab:parameters}
        \begin{tabular}{lcc}
         & \textit{GJ 1214b} & \textit{LHS 1140b} \\
         \hline
        \textit{Stellar Temperature (K)}   & 3500\tnote{*} & 3500\tnote{*}\\
        \textit{Stellar Metallicity}   & 0.29\tnote{1} & -0.15\tnote{3}\\
        \textit{Stellar Gravity}   & 5.0 & 5.0 \\
        \textit{Stellar Radius (R$_{sun}$)}   & 0.215\tnote{2} & 0.215\tnote{3}\\
        \textit{Planetary Mass (M$_{jup}$)}  & 0.0257\tnote{2} & 0.0176\tnote{3}\\
        \textit{Planetary Radius (R$_{jup}$)}  & 0.2446\tnote{2} & 0.1549\tnote{3}\\
        \textit{Equilibrium Temperature (K)} & 600\tnote{2} & 300\tnote{3}\\
    \end{tabular}
\caption{Stellar and planetary parameters used in our atmospheric modeling using current literature values and PICASO grid bounds.}
        \begin{tablenotes}
            \item[*]
            Note that the stellar temperatures are slightly lower, but 3500\,K is the lower limit of the stellar grid used by PICASO.
            \item[1] \citet{Cloutier2021GJ1214b}
            \item[2] \citet{Mahajan2024GJ1214b}
            \item[3] \citet{Cadieux2024lhs1140}
        \end{tablenotes}
    \end{threeparttable}
\end{table}

\section{Error Propagation For Optical Constants}
\label{appendix:errs}
\begin{figure*}[h!]
    \centering
    \includegraphics[width=0.75\textwidth]{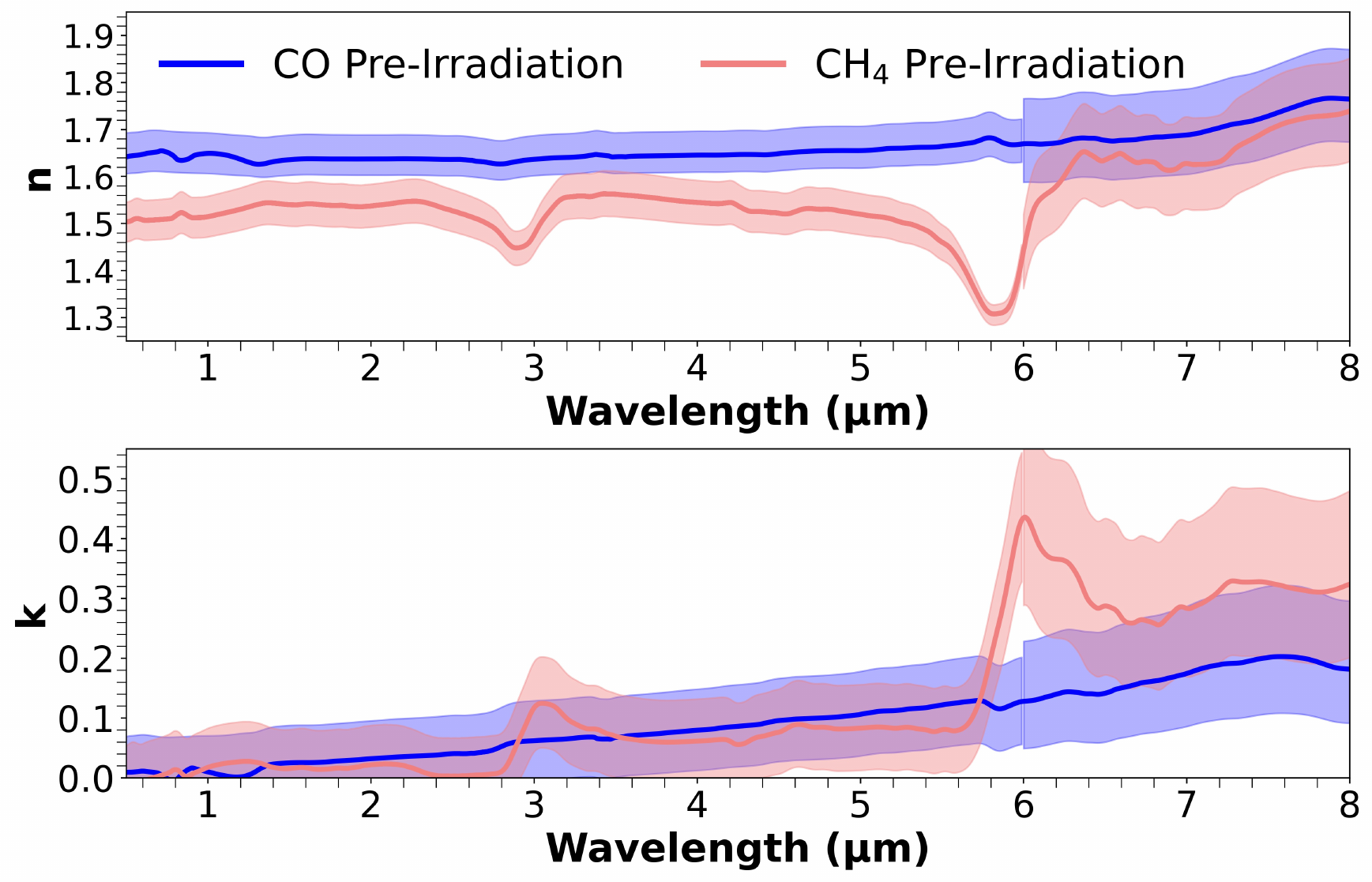}
    \caption{Error Propagation in our $n$ and $k$ values in the visible to mid-IR wavelength region (0.5 to 8 $\mu$m, 24000 to 1250 cm$^{-1}$). Top: Real ($n$) part of the refractive indices and Bottom: imaginary (k) part of the refractive indices. The error bars for the $n$ and $k$ values are derived from propagating error throughout the reflectance and transmission measurements, sample–substrate interface, quantified error in film thickness from \citet{Huseby2025Hazes}, and through Equations \ref{eq:extcoe} and \ref{eq:realref}.}
    \label{fig:errorprop}
\end{figure*}

We note that the $n$ and $k$ values have error due to instrumental setups and constraints, the substrate the haze analog was deposited onto, and the sample–substrate interference in transmittance.
The three-layer structure of the film system, which includes reflection and scattering at the sample–substrate interface, decreases the overall transmittance of our sample, leading to slightly higher $k$ values. 
We have used background subtraction through our instrumentation to eliminate the biases of the instrument setup and substrate, and have added error bars due to the error in film thickness and the sample–substrate interference, specifically from 6 to 8 $\mu$m, where the largest error is found. The post-irradiation values have similar error bars but are not plotted here for clarity.
More in-depth calculations of the error via scattering between the film and substrate would be necessary to fully parameterize the error bars, but this error can be neglected, as these error bars are not large enough to alter the resulting transmission spectrum. 
We urge caution that different instrumental setups, substrates, and thin films introduce the potential to have different results for measured optical constants unless such differences are carefully accounted for. 

\end{document}